# Abnormally High Thermal Conductivity in Fivefold Twinned Diamond Nanowires

Ting Liang[1, 3], Ke Xu[2], Meng Han[3], Yimin Yao[1, 3], Zhisen Zhang[2], Xiaoliang Zeng[3], Jianbin Xu[1, *] and Jianyang Wu[2, 4, **]

[1]Department of Electronic Engineering and Materials Science and Technology Research Center, The Chinese University of Hong Kong, Shatin, N.T., Hong Kong SAR, 999077, P. R. China

[2]Department of Physics, Research Institute for Biomimetics and Soft Matter, Jiujiang Research Institute and Fujian Provincial Key Laboratory for Soft Functional Materials Research, Xiamen University, Xiamen 361005, PR China

[3]Shenzhen Institute of Advanced Electronic Materials, Shenzhen Institutes of Advanced Technology, Chinese Academy of Sciences, Shenzhen 518055, China

[4]NTNU Nanomechanical Lab, Norwegian University of Science and Technology (NTNU), Trondheim 7491, Norway

**Corresponding Authors**

Jianbin Xu[1, *]: *jbxu@ee.cuhk.edu.hk*; Jianyang Wu[1, 4, **]: *jianyang@xmu.edu.cn*

**Abstract**

Fivefold twins (5FTs), discovered nearly 200 years ago, are a common multiply twinned structure that usually dramatically deteriorate the thermal transport properties of nanomaterials. Here, we report the anomalous thermal conductivity ($\kappa$) in a novel fivefold twinned diamond nanowires (5FT-DNWs). The $\kappa$ of 5FT-DNWs is effectively enhanced by the defects of 5FT boundaries, and non-monotonically changes with the cross-sectional area (*S*). Above the critical $S = 7.1$ nm$^2$, 5FT-DNWs show a constant value of $\kappa$, whereas below it, there appears a sharp increase in $\kappa$ with decreasing *S*. More importantly, 5FT-DNWs with minimal *S* show a superior $\kappa$ over the bulk diamond. By confirming the Normal-process-dominated scattering event, it is demonstrated that the phonon hydrodynamic behavior plays a determinative role in abnormally high $\kappa$ of 5FT-DNWs with small *S*. The super-transported phonon hydrodynamic phenomenon unveiled in the twinned diamond nanowires may provide a new route for pursuing highly thermally conductive nanomaterials.

**Keywords:** Fivefold twins; Diamond nanowires; Thermal conductivity; Normal scattering event; Phonon hydrodynamics.

# 1. Introduction

As the cutting-edge micro/nanoelectronic devices/systems are manufactured to be more integrated, miniaturized, and powerful, thermal control and management are becoming increasingly more challenging than ever before[1, 2]. As a result, a charming issue is raising to search for materials with a high thermal power density and good heat dissipation performance[3, 4]. Diamond composed of pure carbon atoms is a promising candidate for removing high-density heat from micro/nanodevices attributed to its ultrahigh room-temperature thermal conductivity ($\kappa$)[5-7], as well as a host of other excellent properties such as chemical inertness[8], high Young's modulus[9, 10], the ultrahigh hardness[11]. However, to adequately meet the heat dissipation demands, the dimensionality of the diamond is required to be reduced down to the nanoscale.

One-dimensional (1D) diamond nanowires (DNWs), as an attractive building block for assembling nanoelectronics and nanophotonics systems, have aroused widespread interest as soon as they were synthesized, because they can function as both nanoscale devices and interconnects[12, 13]. As it is known, reducing the dimensionality of structures to being 1D nanostructures leads to novel transformation in the thermophysical properties[14, 15]. Consequently, there have been a number of theoretical and experimental studies on the thermal transport properties of DNWs. Based on the nonequilibrium molecular dynamics (NEMD) simulations, Jiang[16] and Guo[17] claimed that the $\kappa$ of [110]-oriented DNWs is greater than that of [001] and [111]-oriented DNWs. However, by combining the first principles with the Boltzmann transport equation, Li *et al.*[18] found the [001] growth direction always possesses the largest $\kappa$ in DNWs. Experimentally, Arnoldi *et al.*[19] used laser-assisted atom probe tomography to measure the thermal diffusivity of DNWs for the first time. Unfortunately, the $\kappa$ of all the DNWs considered in the above studies is much lower than that of the bulk diamond, originating from the severe phonon-surface scattering caused by the giant surface-volume ratio in DNWs[15, 20-23]. Thus, it is of scientific and technical interest to find a feasible strategy to enhance the $\kappa$ of DNWs.

Fivefold twins (5FTs) are a common multiply twinned structure that was discovered nearly 200 years ago[24] and have attracted substantial attention in crystal growth[25], biomedical diagnosis[26], optics[27], catalysis[28], and other aspects. In particular, the lattice distortion introduced by 5FT structures can highly improve Young's modulus of the nanowires, which has been repeatedly confirmed in numerous studies[29-32]. A larger Young's modulus means the high elastic constant and group velocity of phonons in the linear dispersion range[33], which is an essential factor in determining $\kappa$. Then, are the 5FT structures helpful to improve the $\kappa$ of DNWs?

In this Letter, the effects of 5FT and the cross-sectional area ($S$) on the thermal transport properties of DNWs are interrogated by extensive molecular dynamics (MD) simulations. It is well known that thermal conductivity $\kappa$ of many different nanowires drops monotonically with decreasing $S$ due to the intensification of boundary scattering[15, 21]. However, we demonstrate that the $\kappa$ of novel fivefold twinned diamond nanowires (5FT-DNWs) varies non-monotonically with the decrease in $S$. Above critical $S = 7.1$ nm$^2$, there exists a negligible change in the $\kappa$, whereas below the critical cross-section, the $\kappa$ of 5FT-DNWs is pronouncedly increased with decreasing $S$, and it is superior to that of the bulk diamond when $S = 1.1$ nm$^2$. The $\kappa$ of DNWs without 5FT structures used as counterparts exhibits the same abnormal non-monotonic behavior. Remarkably, the $\kappa$ of DNWs is effectively enhanced by 5FT, in sharp contrast with the case in other crystals[34-36]. By resolving phonon mean free path (MFP) spectrum and lattice dynamics (including phonon eigenmodes and group velocity), we clarify that the $\kappa$ of DNWs enhanced by 5FT results from the large group velocity and atomic vibrational symmetry. Furthermore, the dissection of the three-phonon scattering rates combined with the phonon dispersion reveals the phonon hydrodynamic behavior in DNWs as the intrinsic physical mechanism responsible for the abnormal enhancement of $\kappa$ at $S < 7.1$ nm$^2$.

## 2. Results and Discussion

Figures 1a and 1b show the simulation geometry model of a typical 5FT-DNW from different view angles. As illustrated, a 5FT-DNW is composed of five DNWs with rhombus-cross sectional geometry, forming a star-shaped 5FT-DNW. The twin boundaries (TBs) in the 5FT-DNWs are highlighted by gray-blue color painting in the hexagonal rings. To indicate the size of 5FT-DNWs, the hexagonal rings in the TBs are marked by digit numbers. Here, 5FT-DNWs with different $S$ are indicated by the number of hexagonal rings, and accordingly, the atomic model in Figure 1a is named "Fivefold_3". Figure 1b shows a perspective view of Fivefold_3, and the heat flux is imposed along the axial $y$-direction in this study. For comparison, TB-free DNWs with the square-shaped cross-section but an identical $S$ to 5FT-DNWs are constructed. For convenience, TB-free DNWs are named based on $S$, e.g., a model with $S = 2.1$ nm$^2$ is straightforwardly named "DNWs_2.1nm$^2$", implying that the TB-free DNWs with different $S$ can be identified by name. More details of the investigated 5FTW-DNWs and TB-free DNWs are given in Supporting Information S1.

We mainly use two MD-based methods with negligible size effects to calculate the $\kappa$ of different DNWs, including the highly efficient homogeneous nonequilibrium MD (HNEMD) method[37, 38] and equilibrium MD (EMD) method based on a Green-Kubo formula[39, 40]. The computational protocols of the two methods and a comparison of their results are shown in Supporting Information S2. Since the two methods are physically equivalent, only the HENMD calculation results are presented in the main text. All the calculation results of $\kappa$ are obtained at room temperature $T = 300$ K.

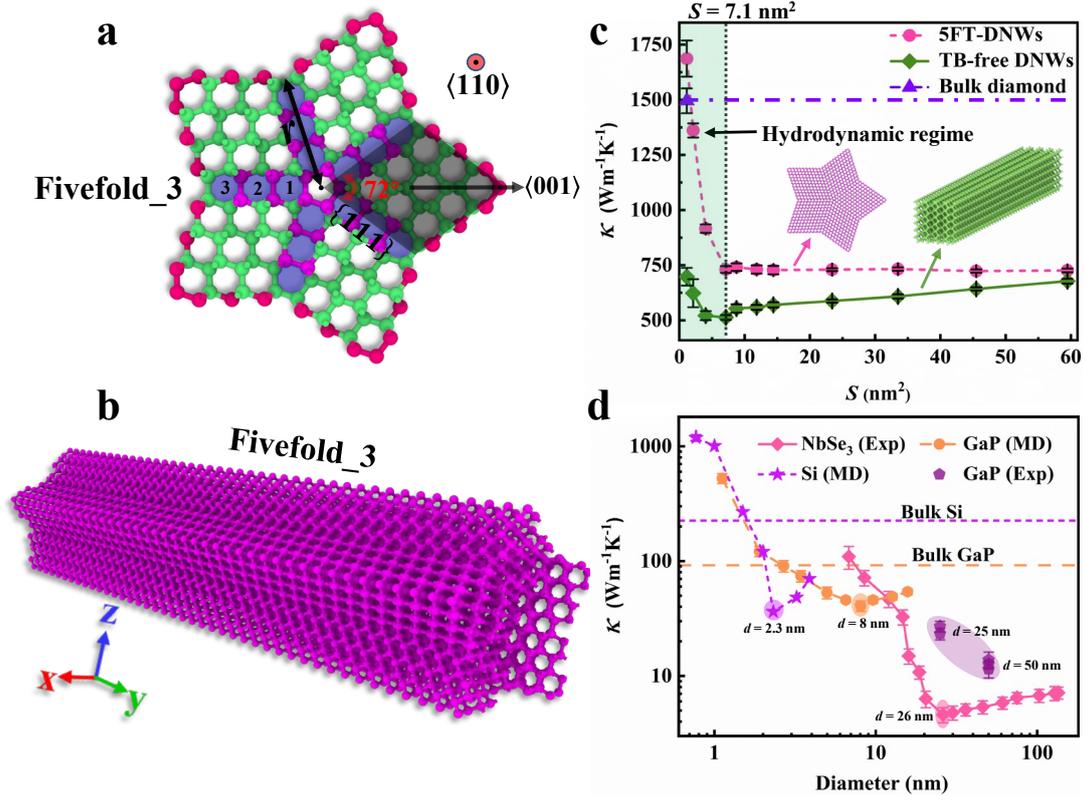

**Figure 1.** (a) Front view for the atomic model of 5FT-DNWs constructed along [110] crystal orientation. (b) Perspective view for the atomic model of 5FT-DNWs (Fivefold_3). (c) The $\kappa$ of different DNWs calculated by the HNEMD method varies with the $S$. The dotted line parallel to the $x$-axis indicates the $\kappa$ of the bulk diamond, while the black dotted line perpendicularly to indicates a paradoxical change in $\kappa$ at $S = 7.1$ nm$^2$. (d) Variation of $\kappa$ (extracted from the literatures) with diameter ($d$) for different nanowire materials, including silicon (Si) nanowires[41], gallium phosphide (GaP) nanowires (results from MD simulation[42] and experimental thermal measurements[43]), and niobium triselenide (NbSe$_3$) nanowires[44]. The short purple underline, and the yellowish dashed line represents the $\kappa$ of the bulk Si[41] and GaP[42], respectively, both calculated from MD. The $\kappa$ of all three nanowires first decreases and then increases rapidly with decreasing $S$. Although the experimental measurements on GaP nanowires are available for only four data points, the $\kappa$ of $d = 25$ nm is greater than that of $d = 50$ nm, demonstrating the same phonon hydrodynamic phenomenon. The experimental measurement lengths of both GaP and NbSe$_3$ nanowires are greater than 10 μm, so the size effect of the obtained $\kappa$ is sufficiently small and can be compared with the results of HNEMD (or EMD).

Figure 1c shows the $\kappa$ of 5FT-DNWs and TB-free DNWs as a function of $S$. Interestingly, both 5FT-DNWs and TB-free DNWs exhibit uniquely non-monotonic changes in $\kappa$ with $S$. For TB-free DNWs, there is a clear crossover in $\kappa$ at the critical $S$ = 7.1 nm$^2$; below which, $\kappa$ increases with decreasing $S$, whereas above which it rises with increasing $S$. With regard to 5FT-DNWs, it is found the intriguing change of $\kappa$ with $S$ that differs from the case of TB-free DNWs. For example, above the critical $S$ = 7.1nm$^2$, $\kappa$ of 5FT-DNWs remains constant, whereas $S$ is below it, $\kappa$ rises sharply as $S$ decreases. More importantly, 5FT-DNWs with $S$ = 1.1 nm$^2$ surprisingly show a larger $\kappa$ than the bulk diamond. Here, the value of $\kappa$ for the bulk diamond is calculated to be 1495.4 ± 56.2 W m$^{-1}$ K$^{-1}$ by HNEMD method, which is lower than the $\kappa$ = 2020 ± 198 W m$^{-1}$ K$^{-1}$[7] measured experimentally by the time-domain thermoreflectance techniques. The underestimation in $\kappa$ for the bulk diamond can be attributed to the excessive anharmonicity of the empirical potentials[45-47] and the quantum effects[48] arising below the Debye temperature[49]. However, in this study, the qualitative comparisons of the MD results based on the same Tersoff potential[50] are rigorous.

In short, the unusual dependences of $\kappa$ on $S$ of 5FT-DNWs and TB-free DNWs are counterintuitive because the surface atoms on smaller nanowires cause stronger phonon boundary scattering[15, 21], resulting in a monotonic reduction in $\kappa$ with decreasing $S$. However, previous studies also reported the abnormal cross-sectional effect on $\kappa$ for other nanowire materials, as shown in Figure 1d, and we will analyze these results in detail later (see the section on phonon hydrodynamic analysis below). Besides, another interest is that, for a given $S$, 5FT-DNWs show a higher $\kappa$ than TB-free DNWs, especially as $S$ < 7.1 nm$^2$. Additional NEMD calculations (Supporting Information S2.3) also reveal that the $\kappa$ of 5FT-DNWs is higher than that of TB-free DNWs for small $S$ in finite sizes. It is generally believed that both nanotwins and 5FT structures break the symmetry of the crystal structure and contribute to stronger anharmonicity phonon scattering[34-36]. Based on this understanding, the presence of 5FT structures should significantly reduce the $\kappa$ of DNWs. Therefore, in the following sections, the physical mechanisms underlying the anomalous changes in $\kappa$ are provided by analyzing three typical systems of 5FT-DNWs and TB-free DNWs with different $S$.

First, the phonon MFP spectrum (see Supporting Information S3.1 for calculation methods) is analyzed to unveil the cross-sectional effect on $\kappa$ and the 5FT-induced enhancement of $\kappa$, since it contains valuable information on the total $\kappa$ based on the classical phonon gas model:

$$\kappa = \frac{1}{3} C_V v_g \lambda, \tag{1}$$

where $C_v$, $v_g$, and $\lambda$ is the volumetric heat capacity, phonon group velocity, and phonon MFP, respectively. Figures 2a and 2b present the MFP spectra $\lambda(\omega)$ and the corresponding length-dependent thermal conductivity $\kappa(L)$ of 5FT-DNWs with three different $S$. It can be clearly seen that $\lambda(\omega)$ decreases with decreasing $S$ for 5FT-DNWs at phonon frequencies $\omega/2\pi < 1$ THz, indicating a stronger phonon-boundary scattering at smaller $S$. The Supporting Information S3.2 provides a more detailed analysis of this using MFP spectrum of the core and surface regions. Phonons with long MFPs at low frequencies can reach the boundaries before full scattering, then this part of phonons will undergo boundary scatterings and cause the drop of MFP. Commonly, stronger phonon scattering leads to smaller $\kappa$. However, the $\kappa(L)$ of 5FT-DNWs with different $S$ breaks this view (see Figure 2b, the structure "Fivefold_3" exhibiting a higher $\kappa$). Similarly, as shown in Figures 2c and 2d, the $\lambda(\omega)$ and $\kappa(L)$ of TB-free DNWs structures demonstrate the same behavior. Unfortunately, the results of MFP spectra $\lambda(\omega)$ do not support the anomalous enhancement of $\kappa$ for DNWs at $S < 7.1$ nm$^2$.

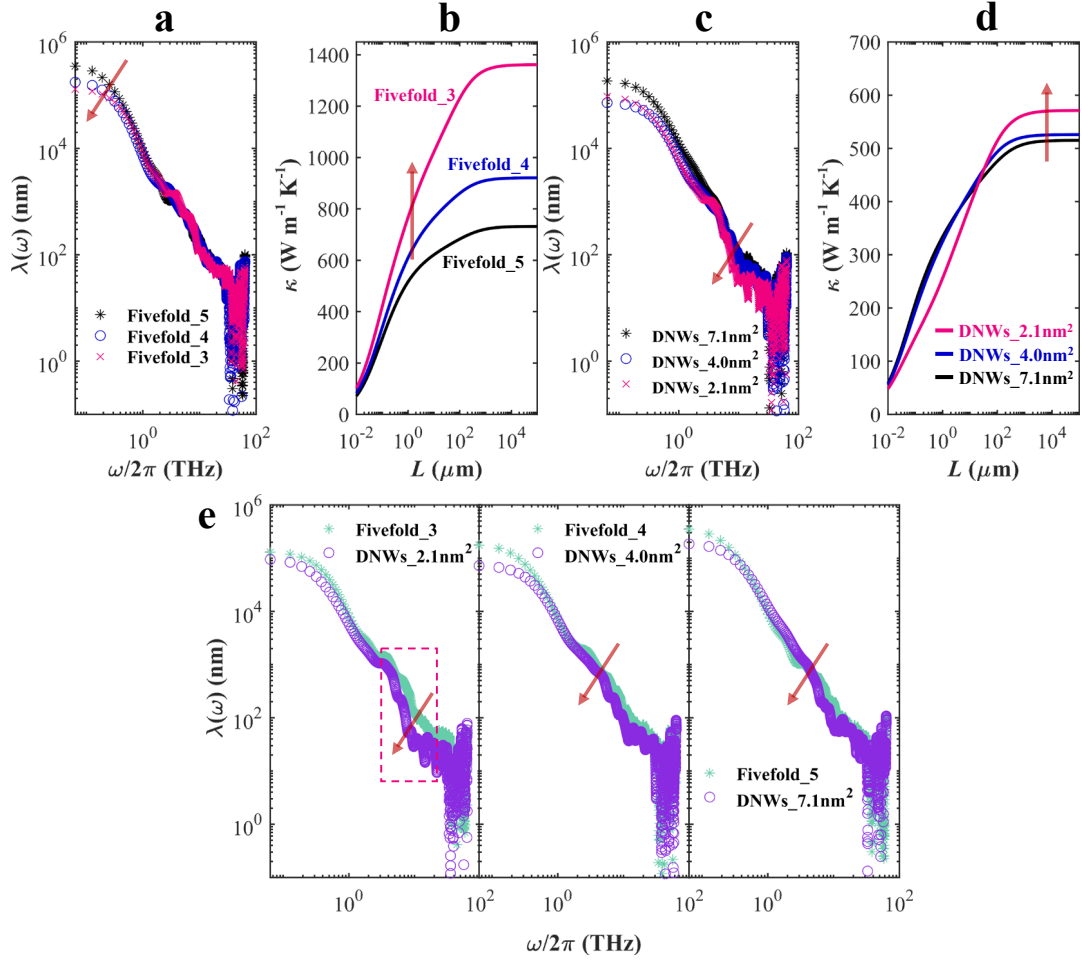

**Figure 2.** The phonon MFP $\lambda(\omega)$ and the length-dependent thermal conductivity $\kappa(L)$ for different DNWs. (a-b) for the 5FT-DNWs, while (c-d) for the TB-free DNWs. (e) A comparison of MFP between 5FT-DNWs and TB-free DNWs at the equivalent $S$. The arrows in the figure serve as an eye guide, and the magenta dashed boxes indicate that phonon eigenvectors in that frequency interval are used for visualization (see Figure 3).

Unexpectedly, as demonstrated in Figure 2e, the MFP of 5FT-DNWs is larger than that of TB-free DNWs especially at low frequencies, despite the extra 5FT-boundary scattering channel appearing in the 5FT-DNWs (see Supporting Information S3.2 for the explanation). Typically, at $S$ = 2.1 nm$^2$, the MFPs of 5FT-DNWs and TB-free DNWs reach 129 and 94 μm at low frequencies, respectively, which is a considerable order of magnitude. This laterally reveals the reasonably large values of $\kappa$ in DNWs with small $S$. In addition, the MFPs of both types of DNWs, including the core and surface regions, are shown in Figure S7, verifying that the $\kappa$ of 5FT-DNWs is higher than that of TB-free DNWs in the whole region. To further understand the mechanism of the long MFP phonons, we visualize the phonon eigenvectors obtained by computing the phonon dispersion (see Supporting Information S3.3 for details). Figure 3 vividly depicts several representative phonon eigenmodes (drawn by OVITO[51]) at Γ point for both types of DNWs.

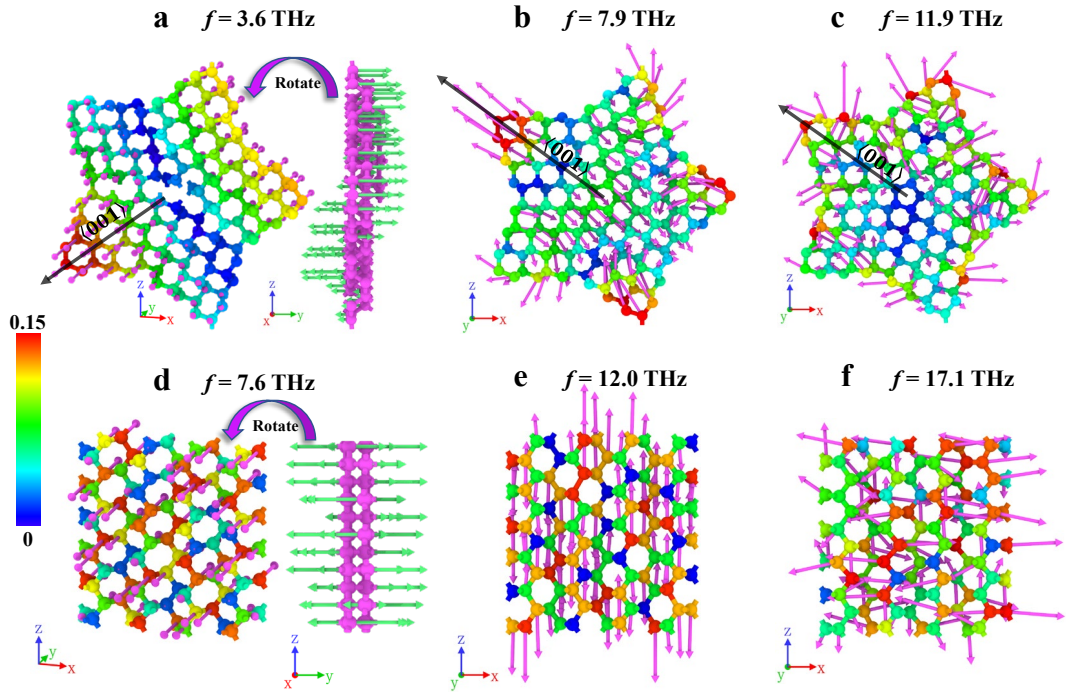

**Figure 3.** The visualization of phonon eigenvectors for two types of DNWs models at representative frequencies. (a–c) for 5FT-DNWs atomic model (Fivefold_3), and (d–f) for TB-free DNWs atomic model (DNWs_2.1nm$^2$). The chromatic bars colored on the atoms represent the amplitude of the normalized eigenvectors. The magnitude and direction of the amplitude are also visualized with arrows. Both (a) and (d) include side and perspective views of the DNWs model to better observe the amplitude and direction of the eigenvectors. The black arrow line in (a–c) represents the equivalent 〈001〉 crystallographic orientation. (a) and (d) represent the propagating phonon modes at low frequencies, and the remaining panels indicate the localized phonon modes.

Figure 3a, d shows the eigenmodes of low-frequency phonons vibrating in a collective movement along the *y*-direction (thermal transport direction), which is the propagating phonon modes that dominate the thermal transport. For the phonon modes

at intermediate frequencies, the vibrations of atoms are in both order and random manners, depending on the local region, as illustrated in Figures 3b and 3e. With increasing frequency, the atomic vibrations become completely disordered and realize into highly localized phonon modes, as depicted in Figures 3c and 3f. Because those localized phonons do not axially propagate through the DNWs and weakly interact with other phonons, their contribution to thermal transport is highly limited[36, 52, 53]. Therefore, it is summarized that the $\kappa$ in both types of DNWs is primarily dominated by the low-frequency phonons, while the contribution of high-frequency phonons to the $\kappa$ can be negligible. More importantly, by visualizing the amplitudes of the eigenvectors, e.g., via comparing Figures 3a and 3d, it can be revealed that the atomic amplitudes of the 5FT-DNWs present a clear symmetry along with equivalent ⟨001⟩ crystallographic orientation, whereas the atomic amplitudes of the TB-free DNWs are fully disordered. Although the amplitude symmetry introduced by the 5FT structures decreases with increasing phonon frequency, it attenuates the disorder and anharmonicity of the atomic vibrations, which leads to an increase in the MFP of the 5FT-DNWs at low frequencies. In terms of the classic phonon gas model, the phonon group velocity is another important determinant of the $\kappa$ in DNWs, and thus it is calculated and shown in Figure S8. By comparison, 5FT-DNWs show larger phonon group velocities than TB-free DNWs at low frequencies, originating from the enhancement of Young's modulus induced by the 5FT structures[29-32]. Additionally, the phonon group velocity of 5FT-DNWs increases with decreasing $S$, particularly in the frequency band less than 20 THz, whereas the phonon group velocity of TB-free DNWs is negligibly changed by the $S$.

In combination with Eq. (1), the physical origins underlying the anomalous thermal transport behavior of DNWs in Figure 1c can be partially unveiled. For 5FT-DNWs with $S > 7.1$ nm$^2$, the constant $\kappa$ is attributed to the competitive balance between the dropping group velocity and the increasing MFP (due to the phonon-boundary scattering suppression). Whereas for TB-free DNWs, the constant phonon group velocity and the increasing MFP eventually yield an increase in $\kappa$ with the enlarging $S$. Comparing the two types of DNWs, the $\kappa$ of 5FT-DNWs is higher than that of TB-free DNWs determined by the relatively large group velocity and the atomic vibrational symmetry introduced by the 5FT structures. Note that the difference in $\kappa$ between the two is more considerable when $S$ is smaller, which is driven by the larger group velocity of 5FT-DNWs at smaller $S$.

The mechanisms underlying the anomalous enhancement of the $\kappa$ of DNWs with $S$ < 7.1 nm$^2$, however, remain unclear. Tracing previous MD calculations and experimental measurements on different nanowires, the same phenomenon of an abnormal increase in $\kappa$ with decreasing diameter $d$ is observed, including Si[41], GaP[42, 43], and NbSe$_3$[44] nanowires (see Figure 1d). Zhou *et al.*[41] have demonstrated a phonon hydrodynamics phenomenon of Si nanowires at small $S$ by probing the three-phonon scattering rates containing Normal (N) and Umklapp (U) processes. The N-process-dominated three-phonon scattering event will lead to the collective movement of phonons and mutual exchange of momentum to form the same drift velocity (e.g., phonon "fluid" flow), which means that the momentum in the system will not decay to zero. In contrast, the U-scattering event will cause the momentum of the phonon modes to equilibrate to zero and obey the Bose-Einstein distribution[41]. Therefore, phonon hydrodynamics occurs when the momentum-conserving N process (which can be considered as a non-resistive scattering event) is much stronger than the momentum-breaking U process (which can be regarded as a resistive scattering event)[41, 54-56].

Inspired by this, we calculate the three-phonon scattering rates (see Supporting Information S3.4 for detailed calculation methods) to explore the contribution of phonon hydrodynamics to anomalous thermal transport. Figure 4 shows the three-phonon scattering rates of 5FT-DNWs and TB-free DNWs with different $S$, while those of the bulk diamond are shown in Figure S9b. As is seen, the scattering rates of the N process is several orders of magnitude stronger than those of the U process in the low-frequency range (< 1 THz) for both 5FT-DNWs and TB-free DNWs with very small $S$, similar to previous studies for Si[41] and GaP[42] nanowires. Typically, for the "Fivefold_2" structure, the scattering rates of the N process can be around 4-orders of magnitude larger than those of the U process in the low-frequency range, as marked by the arrow in Figure 4a. As a result, an evident hydrodynamic character is demonstrated in both types of DNWs, i.e., the formation of hydrodynamic "fluid" flow for the phonons corresponding to the N process. In addition, in Figure 4, the difference between the N- and U-scattering processes of DNWs diminishes as $S$ increases. Therefore, the phonon hydrodynamic behavior formed by the non-resistive N process contributes to an anomalous enhancement of $\kappa$ of the DNWs when $S < 7.1$ nm$^2$. Moreover, besides the large group velocity and lattice vibrational symmetry, the phonon hydrodynamic behavior explains the higher $\kappa$ of the 5FT-DNWs with $S = 1.1$ nm$^2$ than that of the bulk diamond. Because of the weak U-scattering processes in the DNWs system, the $\kappa$ at the minimal $S$ is still convergent[41].

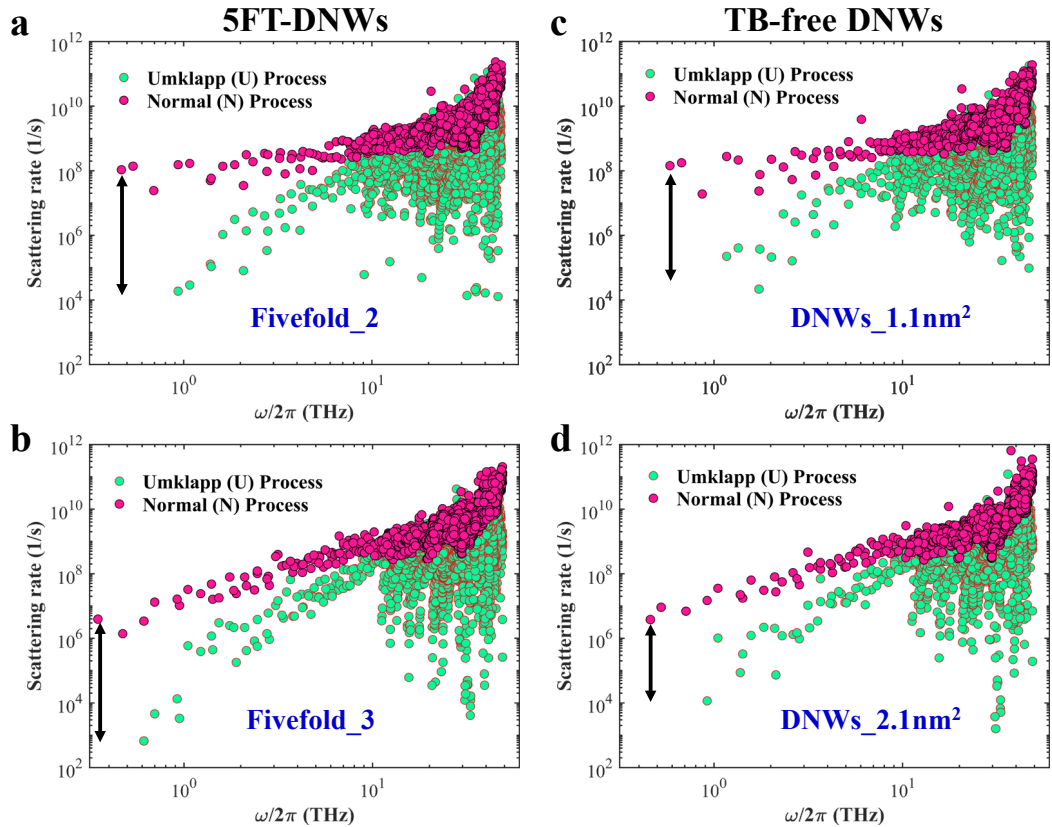

**Figure 4.** Calculated three-phonon scattering rates in two types of DNWs as a function of phonon frequency, including Normal and Umklapp processes: (a–b) for 5FT-DNWs structure, (c–d) for TB-free DNWs structure. The black two-way arrow is a guide for the eyes. Due to the enormous demand for arithmetic power, only the DNWs with minimal $S$ are used for comparison. It should be emphasized that the $S$ is equal for (a) and (c), while the $S$ is equal for (b) and (d).

The suppression of the N-scattering process of DNWs with large $S$ (see Figure 4) can be revealed from the phonon dispersion point of view[41, 42]. Figure 5 shows the phonon dispersion curves of 5FT-DNWs and TB-free DNWs with different $S$. The detailed calculations of the phonon dispersion are described in Supporting Information S3.3. As indicated by the arrows in Figure 5, the optical phonon branches induced by the surface atoms[57, 58] of both types of DNWs are globally downward-shifted as $S$ increases. This signifies that many new scattering channels are activated at low frequencies, for example, with the reaction as: acoustic phonon + acoustic phonon → optical phonon, showing a three-phonon absorption process. As a result, with increasing $S$, more acoustic phonons can be scattered with optical phonons in the low-frequency region where phonon hydrodynamics occurs. Furthermore, because the N-scattering process follows the rule of momentum conservation, those scattering channels prefer the U-scattering process. Therefore, as $S$ is increased from 1.1 nm$^2$, the N process degenerates gradually, making the thermal transport away from the phonon hydrodynamic regime.

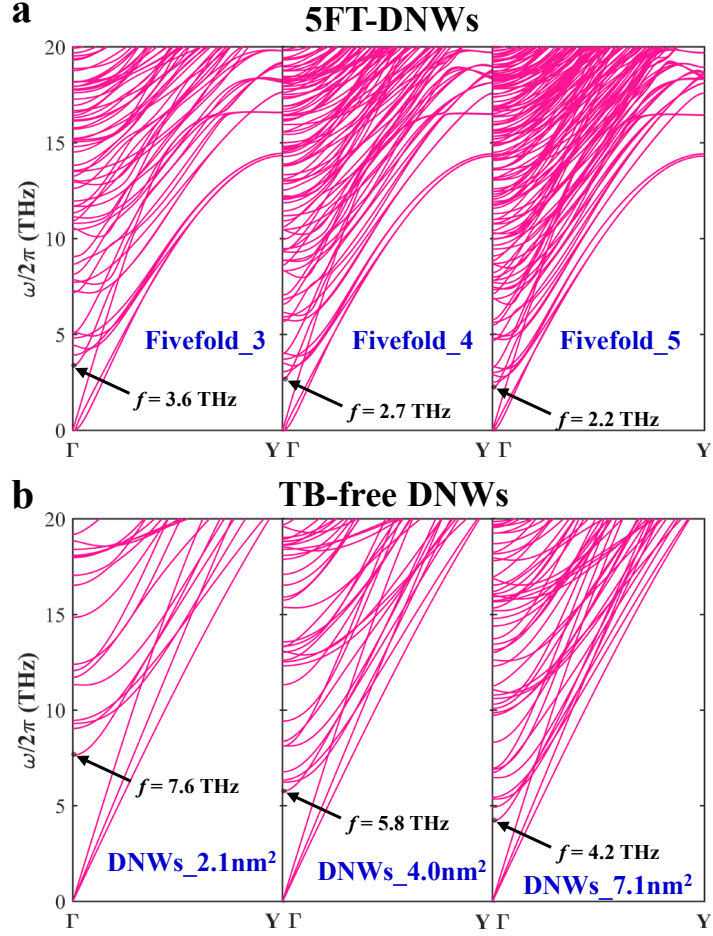

**Figure 5.** Phonon dispersion curves for (a) 5FT-DNWs and (b) TB-free DNWs structures corresponding to different *S*. By considering the presence of a large number of phonon branches in the DNWs, we only visualize phonon dispersion below 20 THz, where low-frequency phonons contribute dominantly to the overall thermal transport of nanowires (as deliberated in the preceding analysis). The black arrow indicates the lowest frequency of the optical phonon branches at the Γ point.

## 3. Conclusions

In summary, by utilizing extensive MD simulations, we find the anomalous $\kappa$ in 5FT-DNWs and TB-free DNWs. For 5FT-DNWs, with decreasing *S*, the $\kappa$ initially remains constant but rises sharply below the critical $S = 7.1$ nm$^2$. For TB-free DNWs, the $\kappa$ first reduces but then increases abnormally below the critical *S*. By comparison, the $\kappa$ of 5FT-DNWs is larger than that of TB-free DNWs. This indicates that the $\kappa$ of DNWs can be effectively enhanced by the defect of 5FT boundaries, in sharp contrast with the case in other crystals. More importantly, 5FT-DNWs with minimal $S$ show a superior $\kappa$ than the bulk diamond. The intrinsic mechanisms of the anomalous phenomena are interrogated by the calculations of phonon MFP, lattice dynamics, and three-phonon scattering rates (containing N- and U-scattering processes). The constant $\kappa$ of 5FT-DNWs with large $S$ stems from a balanced competition between the dropping group velocity and the increasing MFP, whereas the increase in TB-free DNWs with $S > 7.1 nm^2$ is mainly attributed to the rising MFP. Further comparison shows that the atomic vibrational symmetry introduced by the 5FT structures and the relatively large group velocity are responsible for the large $\kappa$ in 5FT-DNWs. Moreover, the phonon hydrodynamic behavior formed by the N-process-dominated phonons is the primary source of anomalous enhancement of the $\kappa$ of DNWs with $S < 7.1 nm^2$. Whereas for DNWs with $S > 7.1 nm^2$, the downward-shift of the surface optical phonons breaks the dominant role of the N-scattering process and causes the thermal transport away from the phonon hydrodynamic regime. This study provides new insights into the thermal transport behavior of nanodiamonds and shows the promise of 5FT-DNWs as an excellent candidate for nanodevice thermal management materials.

## Acknowledgments

T.L. and J. B. X. acknowledge the support from the Research Grants Council of Hong Kong (Grant No. AoE/P-02/12) and the CUHK Group Research Scheme. K.X., J.W., and Z.Z. acknowledge the support from the National Natural Science Foundation of China (Grant Nos. 11772278, 12172314, and 11904300), the Jiangxi Provincial Outstanding Young Talents Program (Grant No. 20192BCBL23029), and the Fundamental Research Funds for the Central Universities (Xiamen University: Grant No. 20720210025). Great thanks to Y. Yu and Z. Xu. of Xiamen University Information and Network Center for their help on high-performance computing.

# Supporting Information for

## "Abnormally High Thermal Conductivity in Fivefold Twinned Diamond Nanowires"

Ting Liang[1, 3], Ke Xu[2], Meng Han[3], Yimin Yao[1, 3], Zhisen Zhang[2], Xiaoliang Zeng[3], Jianbin Xu[1, *] and Jianyang Wu[2, 4, **]

[1]Department of Electronic Engineering and Materials Science and Technology Research Center, The Chinese University of Hong Kong, Shatin, N.T., Hong Kong SAR, 999077, P. R. China

[2]Department of Physics, Research Institute for Biomimetics and Soft Matter, Jiujiang Research Institute and Fujian Provincial Key Laboratory for Soft Functional Materials Research, Xiamen University, Xiamen 361005, PR China

[3]Shenzhen Institute of Advanced Electronic Materials, Shenzhen Institutes of Advanced Technology, Chinese Academy of Sciences, Shenzhen 518055, China

[4]NTNU Nanomechanical Lab, Norwegian University of Science and Technology (NTNU), Trondheim 7491, Norway

**Corresponding Authors**

Jianbin Xu[1, *]: *jbxu@ee.cuhk.edu.hk*; Jianyang Wu[1, 4, **]: *jianyang@xmu.edu.cn*

# Contents

## S1. Atomistic model for different diamond nanowires

In the present work, the novel fivefold twinned diamond nanowires (5FT-DNWs) with a star shape are constructed. Similar to fivefold twinned metallic structures[1-3], a typical star-shaped 5FT-DNW is composed of five identical subunits with [110] axial direction joined along a common quintuple line, as shown in Figure 1a. More colloquially, they consist of five parallel quadrilateral parts (twin domains) with internal angles of 72 degrees arrayed along the nanowire axis in the direction of ⟨110⟩. In Figure 1a, the external boundary atoms appear rose-red, while the fivefold twinned boundaries (TBs) particles are rendered purple-red. Further, the external boundary of each twin domain is located in the {110} facets, and the direction of the fivefold TBs facets between each two twin domains is {111}. The size of 5FT-DNWs is defined by the length ($r$) of fivefold twins, and its cross-sectional area ($S$) can be obtained from the $r$ using the pentagonal area calculation rule.

The square-shaped diamond nanowires (DNWs) without fivefold TBs structures are constructed to compare the calculated results and are termed as TB-free DNWs structures. Since previous computational simulations[4, 5] have demonstrated that [110] crystallographic orientation is the most conductive in DNWs, the model structures of all TB-free DNWs are built along the [110] direction, as shown in Figure S1. We construct the corresponding TB-free DNWs model based on different sizes of 5FT-DNWs, so the cross-sectional areas of the two types of models correspond one-to-one. Although the area of varying size models cannot be exactly the same due to the limitation of lattice parameters, the influence on the calculation of thermal conductivity is minimal. Accordingly, the model in Figure S1 is named "DNWs_2.1nm$^2$", and "2.1nm$^2$" represents the $S$ of the TB-free DNWs.

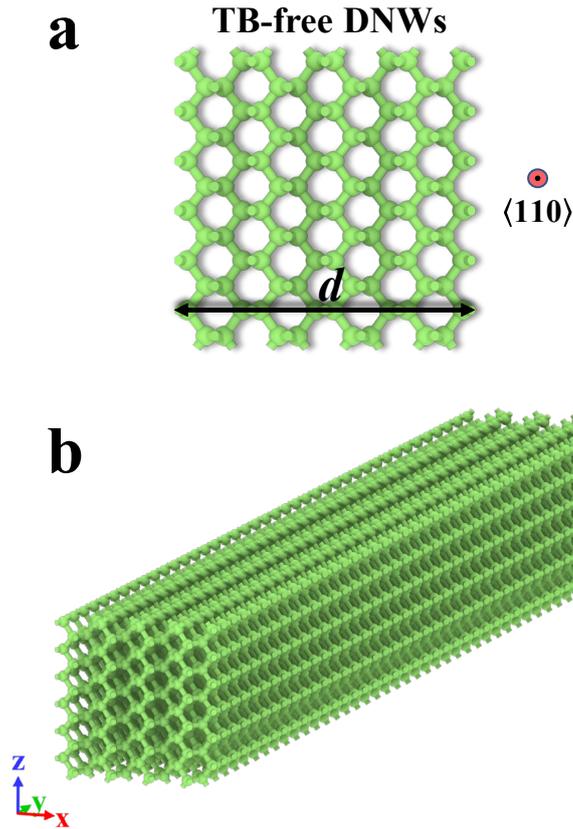

**Figure S1.** The schematic model of TB-free DNWs. (a) Front view for the atomic model of TB-free DNWs constructed along [110] crystallographic orientation. Since the constructed TB-free DNWs model is approximately square, its cross-sectional area is evaluated as $d^2$. (b) Perspective view for the atomic model of TB-free DNWs.

## S2. Methods for thermal conductivity ($\kappa$) calculations

The classical molecular dynamics (MD) simulations are employed to study the thermal transport properties of the 5FT-DNWs and TB-free DNWs under different cross-sectional areas $S$. Since the largest DNWs model studied contains nearly 580,000 atoms, this is the only feasible approach at the fully atomic level. All the MD-based calculations are performed at room temperature (300 K) using the open-source GPUMD (graphics processing units molecular dynamics) package[6-8], in which the standard Newton equations of motion are integrated in time by the velocity-Verlet integration algorithm[9]. The optimized C–C parameters of the Tersoff potential developed in 1989[10] are used to describe the atomic covalent interactions between C, C atoms in the DNWs because they have been proven to be of high quality in evaluating thermal transport properties[4, 11, 12]. In this study, we use different MD-based methods for thermal transport calculations, including the highly efficient homogeneous nonequilibrium MD (HNEMD) method[13, 14] and equilibrium MD (EMD) method based on a Green-Kubo formula[15, 16]. The nonequilibrium MD (NEMD) method[17, 18], which is based directly on Fourier's law of heat conduction, is used only as an auxiliary computational method due to the outrageous low computational efficiency when studying the diffusive transport. Accordingly, we mainly use the HNEMD method and cross-check some results using the other two methods.

The same equilibration procedures are implemented in all three MD methods: the system is first equilibrated at 10 K and zero pressure for 1 ns, and then heated up to 300 K during 1 ns, followed by an NPT equilibration at 300 K and zero pressure for 2 ns and an NVT equilibration at 300 K for 2 ns. The timestep used in the HNEMD calculation is 1 fs (due to the temperature control method used in the data production procedures, resulting in no energy drift[14]), while a timestep of 0.5 fs is used in both the EMD and NEMD calculations. Rigorous tests have shown that using different timesteps does not compromise the results while ensuring the energy conservation of the systems. The effects of temperature and external pressure are not considered here. Periodic boundary conditions are used in the transport direction ($y$-direction) in HNEMD and EMD methods, and conversely, fixed boundary conditions are used in NEMD simulations. Perpendicular to the heat transport direction ($x$ and $z$ directions), all three methods employ free boundary conditions to ensure that the models can be fully optimized. In the following, we will describe the different MD approaches and their data production procedures.

**S2.1. The HNEMD method**

Based on the non-canonical linear-response theory, the HNEMD method was first proposed by Evans *et al.*[13] in terms of two-body potentials. Recently, this method was extended to general many-body potentials[14] and integrated into the GPUMD

package. In this method, after equilibration (as mentioned above), an external force:

$$\vec{F}_i^{\text{ext}} = E_i \vec{F}_e + \sum_{j \neq i} \left( \frac{\partial U_j}{\partial \vec{r}_{ji}} \otimes \vec{r}_{ij} \right) \cdot \vec{F}_e \tag{S1}$$

is exerted to each atom $i$, driving the system out of equilibrium. Driven by $\vec{F}_e$ (of dimension inverse length), the "hotter" atoms are pushed in the direction of the transport ($y$ direction, see Figure 1b and Figure S1b), and the "cooler" atoms are pulled in the opposite direction, while the overall temperature of the system is maintained near the target temperature (300 K) using a Nosé-Hoover thermostat[19, 20]. Then, an ensemble average steady-state nonequilibrium heat current $\langle J(t) \rangle_{ne}$, which is proportional to the magnitude of the parameter $\vec{F}_e$, will be generated. Finally, the thermal conductivity $\kappa$ along heat transport direction is given by:

$$\kappa(t) = \frac{1}{t} \int_0^t \frac{\langle J(t) \rangle_{ne}}{TVF_e} dt, \tag{S2}$$

where $t$, $V$, and $T$ are the production time, system volume, and temperature. From the multiple independent values at $t = 20$ ns, we obtain the $\kappa$ values and their error estimates (standard deviation divided by the square root of the number of independent runs) for different DNWs structures. The results of HNEMD calculations for two typical DNWs with the same $S$ are shown in Figure S2. The $\kappa$ of the structure "Fivefold_3" is larger than that of the structure "DNWs_2.1nm$^2$".

Note that the driving force parameter $\vec{F}_e$ needs to be chosen carefully to ensure a large data signal-to-noise ratio within the linear response of the systems[14]. A rule of thumb[21] states that when $\vec{F}_e \lambda_{\max} \lesssim 1$, where $\lambda_{\max}$ denotes the maximum phonon mean free path (MFP), the linear response is wholly guaranteed. In this work, we conduct strict tests on the $\vec{F}_e$ used in all the DNWs systems and ensure that the running thermal conductivity data obtained is neither divergent but also has a good signal-to-noise ratio. The specific data are listed in Table S1. Since there are significant differences in the order of magnitude of convergent $\kappa$ for different structures calculated by HNEMD, the $\vec{F}_e$ used varies and the number of independent runs also varies.

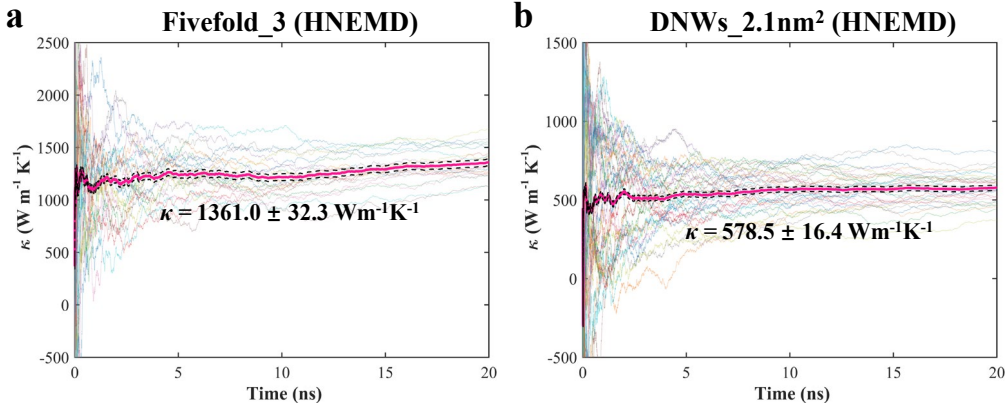

**Figure S2.** Cumulative average of the thermal conductivity as a function of time for different DNWs structures from the HNEMD simulations. In each panel, the thin transparent lines are from multiple independent runs (see Table S1), and the thick magenta solid and black dashed lines represent the average and error bounds from the individual runs.

### S2.2. The EMD method

We use the EMD method to cross-check the HNEMD results for different DNWs structures. As shown below, the running thermal conductivity $\kappa(\tau)$ can be deduced as an integral of the heat current autocorrelation function $\langle J_y(0)J_y(\tau)\rangle$ (taken the y-direction as the heat transport direction) with respect to the correlation time $\tau$

$$\kappa(\tau) = \frac{1}{k_\mathrm{B}T^2V}\int_0^\tau \langle J_y(0)J_y(t)\rangle dt. \tag{S3}$$

Here, $k_\mathrm{B}$ is Boltzmann's constant, $T$ is the system temperature, and $V$ is the system volume. After achieving equilibrium using the procedure described above, the heat flow is sampled at equilibrium states for 10 ns (in the NVE ensemble), and the Green-Kubo formula is used to calculate the thermal conductivity as a function of the correlation time. The results of the running thermal conductivity $\kappa(\tau)$ calculated by EMD for the typical structures are shown in Figure S3, and the final $\kappa$ values are obtained by averaging the results for the last 400 ps. Due to the differences in convergent $\kappa$ of different structures, we performed 50 to 100 independent runs (listed in Table S1) and calculated the error bounds in terms of the standard error.

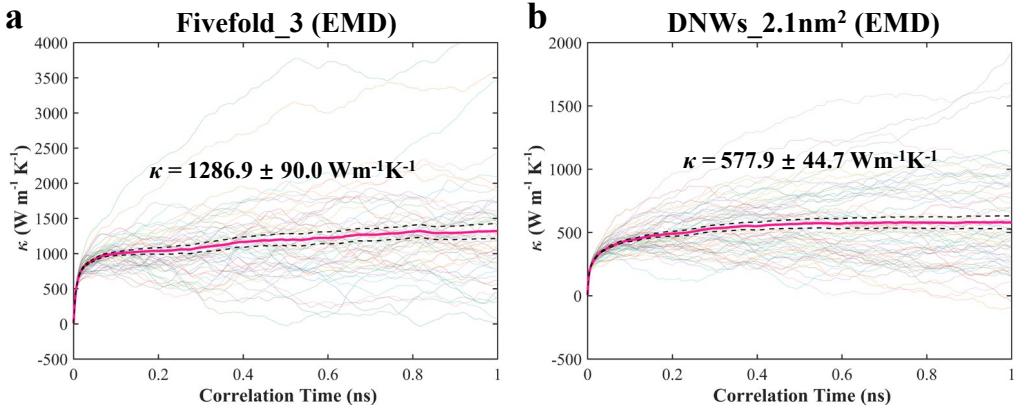

**Figure S3.** Running thermal conductivity $\kappa(\tau)$ for different DNWs structures at 300 K and zero pressure as a function of correlation time. The meanings of the different line shapes are the same as in Figure S2.

By comparing the data in Table S1, we find that the thermal conductivities obtained for the same structure in both HNEMD and EMD calculations are equal within the standard error, which indicates that the two methods are physically equivalent[14]. Note that the total production time of HNEMD simulation is shorter and the statistical error is smaller than that of EMD simulation, reflecting that the HNEMD method is more efficient than the EMD method. So, we only use the EMD method to calculate the thermal conductivity of some DNWs structures. The cross-validation reflects that our thermal conductivity results are self-consistent and credible.

**Table S1.** Relevant data from the HNEMD and EMD simulations. $M$ is the number of independent runs, while $N$ indicates the number of atoms the system contains. $\kappa_{ave}$ (in units of W m$^{-1}$ K$^{-1}$) is the average thermal conductivity and its standard error is $\kappa_{err}$. The unit of $F_e$ is $\mu m^{-1}$, and the unit of cross-sectional area is nm$^2$. The simulation cell length for different DNWs structures in both EMD and HNEMD simulations is 50 nm, which ensures that the thermal conductivity is not affected by the length of the system (negligible finite-size effect). The black bolded in the table indicates that the energy of the system is not conserved under EMD calculations, a behavior that also occurs in silicon nanowires[22].

| Area | Structure | N | HNEMD | | | | EMD | | |
|---|---|---|---|---|---|---|---|---|---|
| | | | M | $F_e$ | $\kappa_{ave}$ | $\kappa_{err}$ | M | $\kappa_{ave}$ | $\kappa_{err}$ |
| \ | Bulk_diamond | 4096 | 15 | 0.05 | 1495.4 | 56.2 | 50 | 1481.8 | 99.2 |
| 1.1 | Fivefold_2 | 16745 | 40 | 0.02 | 1686.7 | 82.2 | 100 | 1611.0 | 75.4 |
| | DNWs_1.1nm$^2$ | 9408 | 35 | 0.03 | 697.7 | 40.4 | 60 | **770.8** | **44.8** |
| 2.1 | Fivefold_3 | 30845 | 20 | 0.03 | 1361.0 | 32.3 | 60 | 1286.9 | 90.0 |
| | DNWs_2.1nm$^2$ | 18816 | 20 | 0.03 | 578.5 | 16.4 | 60 | 577.9 | 44.7 |
| 4.0 | Fivefold_4 | 48755 | 10 | 0.03 | 915.9 | 19.2 | 50 | 853.1 | 65.5 |
| | DNWs_4.0nm$^2$ | 37632 | 10 | 0.03 | 529.0 | 15.5 | 50 | 529.4 | 40.4 |
| 7.1 | Fivefold_5 | 70645 | 10 | 0.03 | 730.9 | 15.5 | 50 | 696.8 | 53.5 |
| | DNWs_7.1nm$^2$ | 60368 | 10 | 0.03 | 513.3 | 13.4 | 50 | 497.5 | 45.8 |
| 8.7 | Fivefold_6 | 96515 | 10 | 0.03 | 741.2 | 17.2 | 50 | 698.1 | 60.9 |
| | DNWs_8.7nm$^2$ | 75264 | 10 | 0.05 | 555.6 | 20.4 | 50 | 532.4 | 49.6 |
| 11.8 | Fivefold_7 | 100965 | 10 | 0.05 | 730.8 | 18.8 | 50 | 735.9 | 51.7 |
| | DNWs_11.8nm$^2$ | 87920 | 10 | 0.05 | 559.0 | 11.1 | 50 | 557.6 | 38.6 |
| 14.4 | Fivefold_8 | 160195 | 10 | 0.05 | 727.9 | 19.2 | 50 | 721.3 | 56.9 |
| | DNWs_14.4nm$^2$ | 110528 | 10 | 0.05 | 570.3 | 11.9 | 50 | 574.3 | 35.9 |

## S2.3. The NEMD method

In finite size, we use the NEMD method to compare the thermal conductivity of 5FT-DNWs with TB-free DNWs because it is an inhomogeneous method that requires a heat source and a sink. A Langevin thermostat with a coupling time of 0.1 ps is used to generate the heat source and sink according to the recommendation of reference[23, 24]. The apparent thermal conductivity in the transport direction ($y$-direction, see Figure S4a) is calculated as[23]:

$$\kappa(L) = \frac{dE/dt}{S\Delta T/L}, \tag{S4}$$

where $S$ is the cross-sectional area in the transverse directions and $dE/dt$ is the average energy exchange rate between the thermostats and the thermostatted regions. In the

NEMD method, we only calculate the thermal conductivity of the DNWs at a system length $L$ of 50 nm. The results of NEMD are shown in Figure S4b, which are consistent with those of HNEMD and EMD, i.e., the thermal conductivity of the 5FT-DNWs structure is larger than that of TB-free DNWs in the finite size. In addition, a comparison about the NEMD data with the length-dependent thermal conductivity $\kappa(L)$ (see S3.1 for calculation methods) is carried out in Figure S4c, and it can be seen that the thermal conductivity of NEMD agrees well with that of $\kappa(L)$ at $L = 50$ nm. The thermal conductivity of structure "DNWs_7.1nm$^2$" is slightly higher than that of structure "DNWs_4.0nm$^2$" at the finite size. It is worth mentioning that the $\kappa(L)$ of the DNWs structures only fully converges when $L \approx 1$ mm, which means that the non-monotonic dependence of $\kappa$ on $S$ and the phonon hydrodynamic behavior in the main text are hardly captured by the finite-size NEMD method.

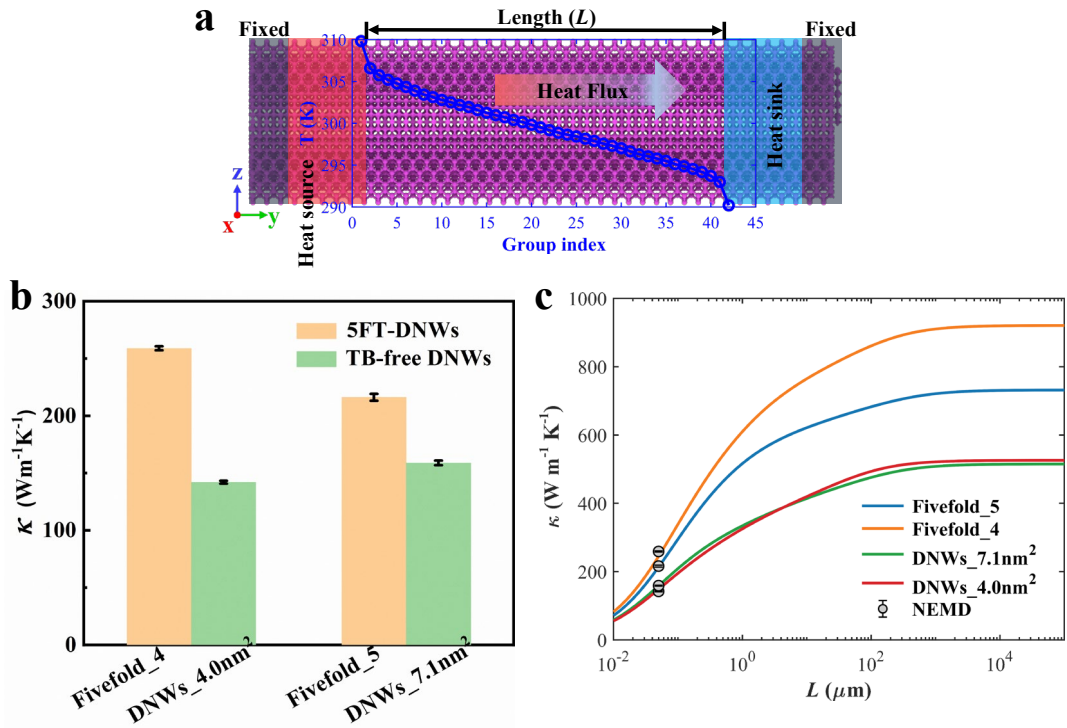

**Figure S4.** (a) Schematic illustration of the setups for the NEMD simulations. The blue inset shows the temperature profile at a steady state. The effective length $L$ between the heat source and the heat sink is divided into 40 groupings. (b) The thermal conductivity of different DNWs calculated by the NEMD method at 300 K. For each simulation system, we perform four

independent runs and estimate error bounds based on standard deviations. (c) A comparison of the length-dependent thermal conductivity κ(L) of four typical DNWs with NEMD data.

## S3. Calculations of phonon properties

### S3.1. Spectral heat current decomposition

To analyze the thermal conductivity results more quantitatively, the spectral heat current (SHC) decomposition method[14, 25-27] is conducted to obtain the phonon frequency-dependent MFP and the length-dependent thermal conductivity κ(L). In this approach, the spectral conductance $G(\omega)$ in the ballistic regime is first calculated using NEMD simulations, then the spectral thermal conductivity $\kappa(\omega)$ in the diffusive regime is given using HNEMD simulations. Accordingly, the spectral phonon MFP is determined by:

$$\lambda(\omega) = \frac{\kappa(\omega)}{G(\omega)}. \tag{S5}$$

The above calculations of $G(\omega)$ and $\kappa(\omega)$ involves the force-velocity correlation function evaluated in the nonequilibrium steady state given by the following [27]:

$$\vec{K}(t) = \sum_i \langle \mathbf{W}_i(0) \cdot \vec{v}_i(t) \rangle, \tag{S6}$$

where $\mathbf{W}_i$ and $\vec{v}_i$ denote the virial tensor and velocity of atom $i$, respectively. Then one can calculate the $G(\omega)$ and $\kappa(\omega)$ from the following Fourier transform $\tilde{\vec{K}}(\omega)$ of $\vec{K}(t)$:

$$G(\omega) = \frac{2\tilde{\vec{K}}(\omega)}{V \Delta T}, \quad \kappa(\omega) = \frac{2\tilde{\vec{K}}(\omega)}{VTF_\text{e}}. \tag{S7}$$

From the spectral decomposition and the first-order extrapolation of the standard length scaling formula for the thermal conductivity[28],

$$\frac{1}{\kappa(L)} = \frac{1}{\kappa_\infty}\left(1 + \frac{\lambda}{L}\right), \quad (\kappa_\infty = \kappa_{\text{HNEMD}}). \tag{S8}$$

We can obtain the length-dependent thermal conductivity $\kappa(L)$:

$$\kappa(L) = \int_0^\infty \frac{d\omega}{2\pi} \kappa(\omega, L), \tag{S9}$$

where

$$\frac{1}{\kappa(\omega, L)} = \frac{1}{\kappa(\omega)}\left(1 + \frac{\lambda(\omega)}{L}\right). \tag{S10}$$

Based on Eq. (S7), the HNEMD-based and NEMD-based SHC results can be obtained. Meanwhile, the phonon frequency-dependent MFP and length-dependent thermal conductivity $\kappa(L)$ of different DNWs structures can be accessed by Eqs. (S5) and (S9), respectively.

**S3.2. Phonon MFP spectrum of core and surface regions**

To further evaluate the effect of TBs on phonon scattering in DNWs, we calculate the phonon MFP spectrum for different regions of DNWs. Figure S5 shows the definition of the surface and core regions in both types of DNWs. The protocols for calculating the phonon MFP spectrum in the different regions are as follows. We first calculated the spectral thermal conductance $G(\omega)$ and spectral thermal conductivity $\kappa(\omega)$ of the surface/core regions using NEMD and HNEMD, respectively, and then obtained the spectral phonon MFP of desired regions using Eq. (S5). Finally, using Eq. (S9), the corresponding length-dependent thermal conductivity $\kappa(L)$ of different regions can be extracted. Notice that we treat the model as a complete whole and impose the corresponding heat fluxes, dividing only the surface/core regions into different groups for phonon MFP extraction. In addition, the area of the core region will remain constant for both different types of DNWs, which means that the proportion of surface region increases as the total cross-sectional area $S$ increases. Such an arrangement can better help to understand the surface phonon scattering and the $\kappa$ of different regions with different *S*.

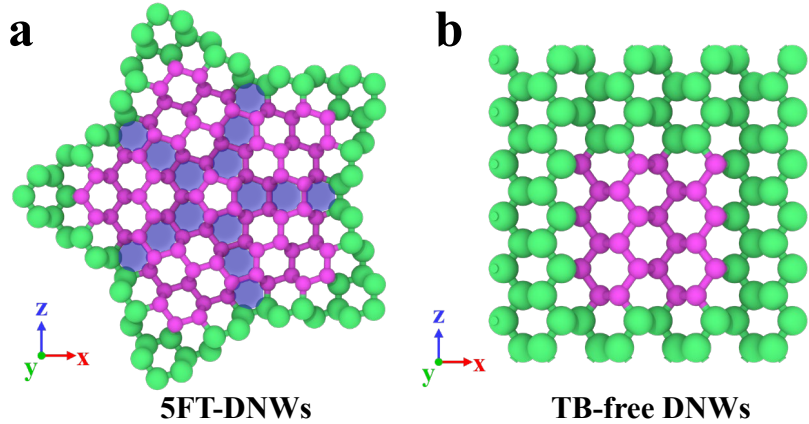

**Figure S5.** A schematic diagram of the surface/core region of different DNWs selected to calculate the phonon MFP spectrum. (a) and (b) are snapshots of the typical structures of 5FT-DNWs and TB-free DNWs, respectively. The smaller diameter purple atoms represent the core region, while the thicker green atoms indicate the surface region.

Figures S6a and S6b compare the MFP spectrum of the two types of DNWs in the core and surface regions. The comparisons show intuitively that as the *S* decreases, the phonon MFP ($\omega/2\pi$ < 1 THz) of both nanowire structures decreases to a greater extent in the core region than in the surface region. This confirms that as *S* gets smaller, the core region of the DNWs (including 5FT-DNWs and TB-free DNWs) is more severely affected by surface effects, which also means that the phonon-boundary scattering is more intense. Interestingly, the phonon MFP in the surface region of TB-free DNWs is nearly unchanged, while the MFP in the surface region of 5FT-DNWs still decreases with *S* decreasing. This indicates that the fivefold twinned-boundary scattering is present in the 5FT-DNWs structure (see Figure 1a) due to the increasing proportion of fivefold TBs as the area of the surface region decreases. We note that at small *S* (e.g., "Fivefold_3" and "DNWs_2.1nm$^2$"), the MPF of both types of DNWs increases at high frequencies (> 60 THz), but the contribution to the $\kappa$ is minimal and is therefore neglected. Correspondingly, the length-dependent thermal conductivity for different regions of different structures is calculated as depicted in Figures S6c and S6d. In both types of DNWs, as $S$ decreases, the $\kappa(L)$ of surface region *vs.* $\kappa(L)$ of the core region becomes increasingly dominant (the ratio $\kappa_{core}(L)/\kappa_{surface}(L)$ of both undergoes a shift from < 1 to > 1), which further suggests that phonon-boundary scattering becomes more severe at small $S$.

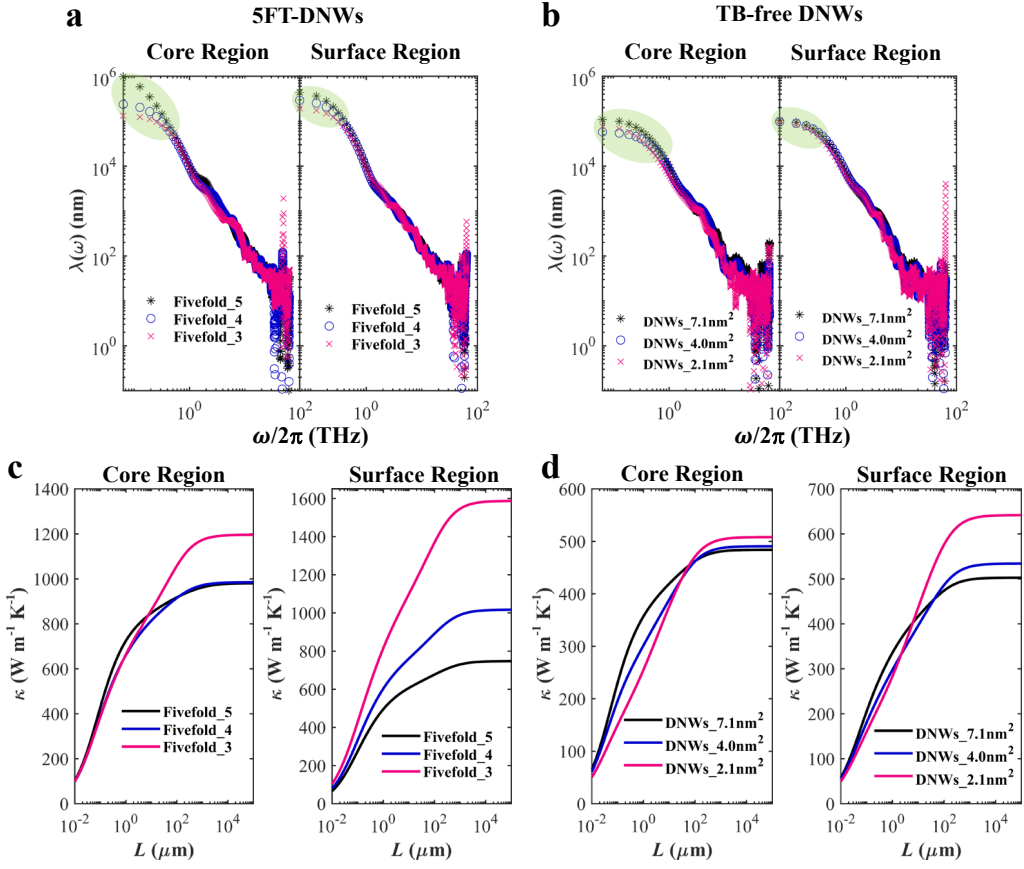

**Figure S6.** A comparison of the MFP spectrum $\lambda(\omega)$ and the length-dependent thermal conductivity $\kappa(L)$ of different regions in the same DNWs. (a) and (c) for comparing 5FT-DNWs type structures, while (b) and (d) for comparing TB-free DNWs type structures. The light green shaded part highlights the comparison of the MFP spectrum.

To investigate the physical mechanism of the higher thermal conductivity (see Figure 1) of 5FT-DNWs compared to that of TB-free DNWs, the MPFs of different sub-regions are also depicted in Figure S7. Obviously, the MFP of 5FT-DNWs is larger than that of TB-free DNWs in both the core and surface regions, although the fivefold twinned-boundary scattering exists in 5FT-DNWs. This is attributed to the symmetry of the lattice vibrations introduced by the 5FTs structure, as explained in the main text.

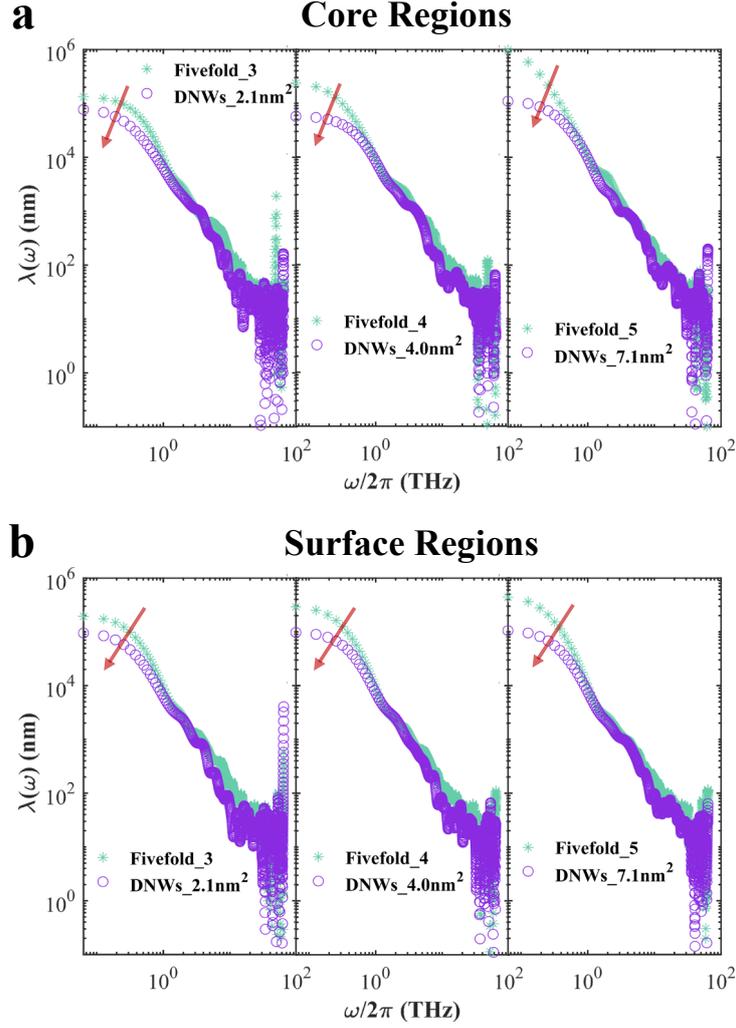

**Figure S7.** A comparison of MFPs of different types of DNWs for the core (a) and surface (b) regions. The arrows in the figure serve as an eye guide.

### S3.3. The phonon dispersion and group velocity

To obtain more phonon information, we calculate the phonon dispersion diagram of the two types of DNWs using lattice dynamics methods and then extract the corresponding group velocities and eigenvectors. Atomic forces are collected by the LAMMPS package[29] using the empirical Tersoff potential[10]. The second-order interatomic force constants (IFCs) are obtained using the finite difference method with an atomic displacement of 0.01 Å. The second-order IFCs are then fed to the PHONOPY package[30] to construct the dynamical matrix, and subsequently, the phonon frequencies are calculated as a function of the wave vector $q$ (i.e., the phonon dispersion relations and the eigenvectors also obtained). Finally, the group velocities can be accessed by making a first-order derivative of the phonon dispersion using the finite difference method, as shown in Figure S8. In both DNWs, a $1 \times 3 \times 1$ supercell and a $1 \times 101 \times 1$ $q$-point grid are used to sample the first Brillouin zone, which provides sufficient accuracy for phonon calculations. To verify the plausibility of the phonon dispersion of DNWs, the phonon dispersion of the bulk diamond is calculated and visualized in Figure S9a. Our result agrees very well with that in Ref.[31].

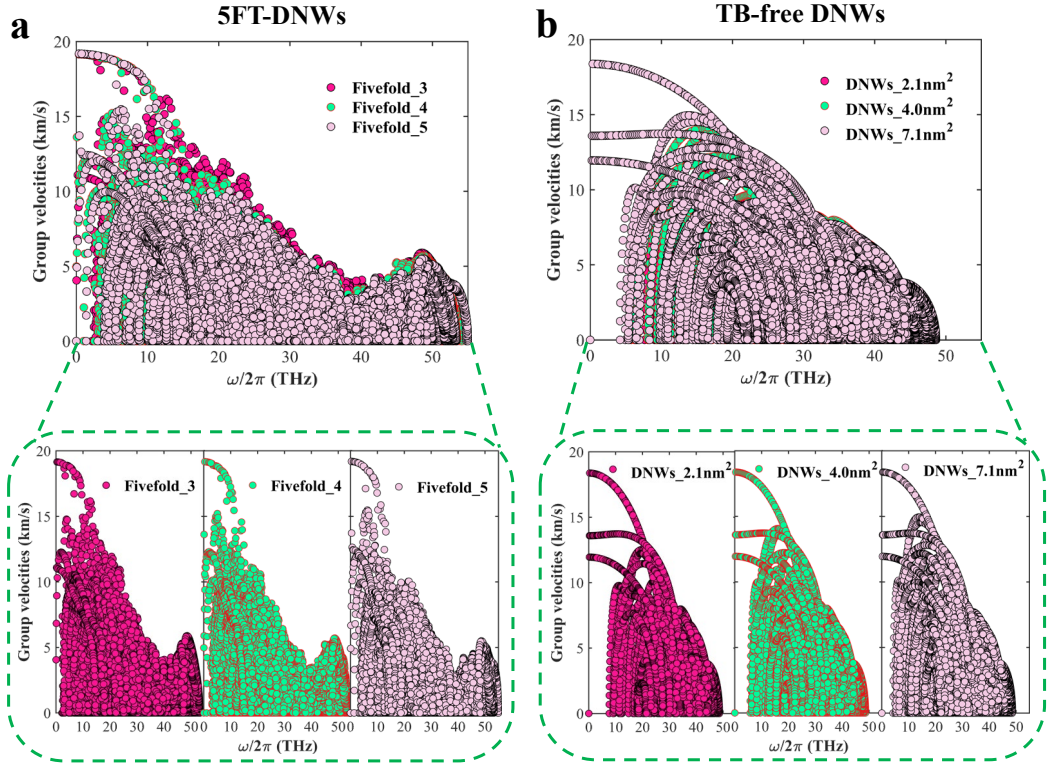

**Figure S8.** Phonon group velocities as a function of phonon frequencies for (a) 5FT-DNWs and (b) TB-free DNWs. Since the group velocities of different $S$ in TB-free DNWs have slight variation, they are covered by each other at low frequencies (< 10 THz) and high frequencies (> 20 THz), resulting in the eye not being able to distinguish them from the diagram. To better show the physical picture of the phonon group velocities, they are unwrapped in the local images.

### S3.4. The three-phonon scattering rates

To clarify the phonon hydrodynamic behavior in 5FT-DNWs and TB-free DNWs structures, we calculate the three-phonon scattering rate at 300 K using

PHONO3PY[32], which includes the Normal process and Umklapp process. The third-order IFCs, also computed using the finite difference method with an atomic displacement of 0.03 Å, are used to obtain the imaginary part of the phonon self-energy $\Gamma_\lambda(\omega)$ and to extract the phonon lifetime, which is calculated by the single-mode relaxation time approximation (a method for solving the Boltzmann transport equation). For the bulk diamond, a 3 × 3 × 3 supercell and a 30 × 30 × 30 $q$-point grid are used to sample the first Brillouin zone, and for both types of DNWs, a 1 × 3 × 1 supercell and a 1 × 150 × 1 $q$-point grid are used. Using such a dense grid is sufficient to obtain accurate three-phonon scattering rate results. The summations over the first Brillouin zone are done using the linear tetrahedron method[33, 34]. Since the calculation of the $q$-point triplet integrals of $\Gamma_\lambda(\omega)$ is performed separately for the Normal $\Gamma_\lambda^N(\omega_\lambda)$ and Umklapp $\Gamma_\lambda^U(\omega_\lambda)$ processes (e.g., $\Gamma_\lambda(\omega_\lambda) = \Gamma_\lambda^N(\omega_\lambda) + \Gamma_\lambda^U(\omega_\lambda)$), we can easily capture the scattering rates of the Normal and Umklapp processes. For more detailed calculation details, one can refer to the PHONO3PY usage tutorial[35]. Due to the vast computational resources required to calculate three-phonon scattering, we cannot calculate DNWs structures with $S$ exceeding 2.1 nm$^2$. We calculate the three-phonon scattering rates for bulk diamond, as shown in Figure S9b. At low frequencies, the difference between N and U processes in bulk diamond is within an order of magnitude[36], which is smaller than that in DNWs (see Figure 4 in the main text).

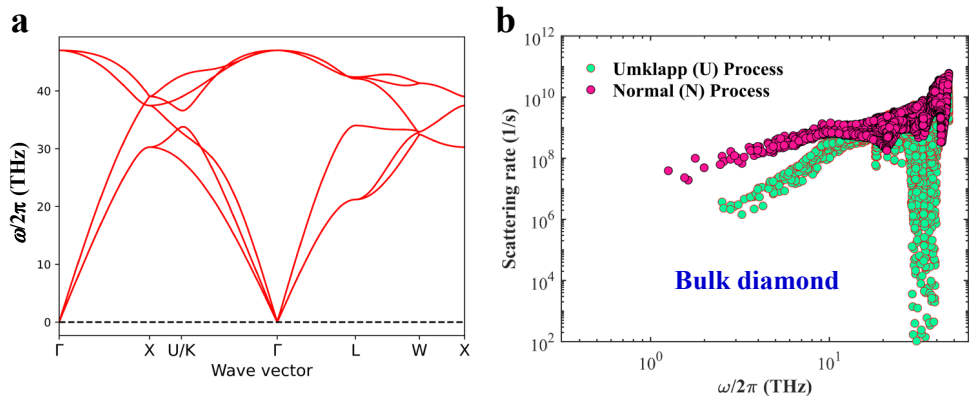

**Figure S9.** (a) The phonon dispersion curve and (b) the three-phonon scattering rates for bulk diamond.

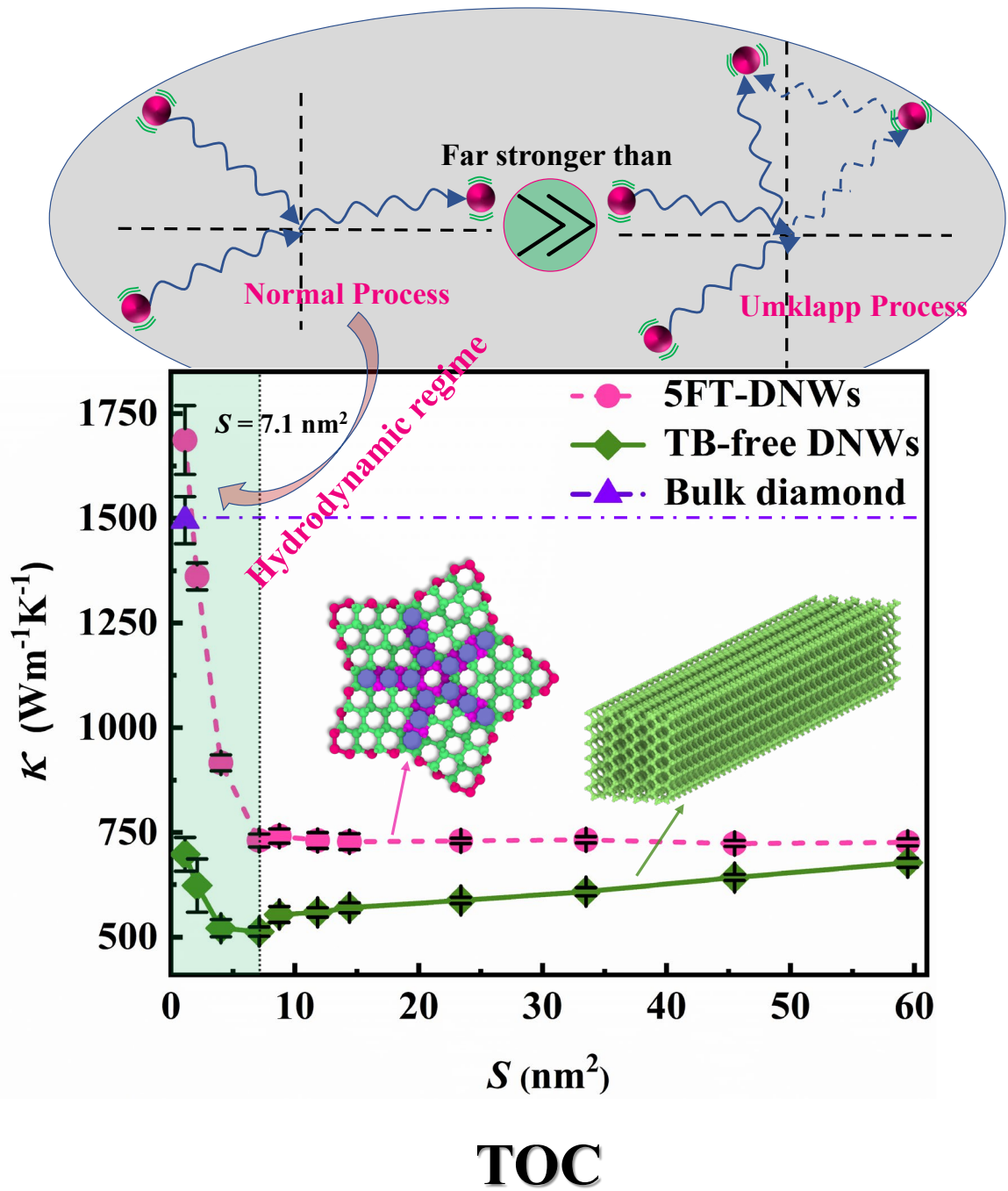

TOC